\documentclass[fleqn,usenatbib]{mnras}

\usepackage{newtxtext,newtxmath}

\usepackage[T1]{fontenc}

\DeclareRobustCommand{\VAN}[3]{#2}
\let\VANthebibliography\thebibliography
\def\thebibliography{\DeclareRobustCommand{\VAN}[3]{##3}\VANthebibliography}

\usepackage{graphicx}	
\usepackage{amsmath}	
\usepackage{xcolor}
\usepackage[export]{adjustbox}
\usepackage{flushend}
\usepackage[utf8]{inputenc}
\usepackage[normalem]{ulem}
\usepackage{comment}
\usepackage{url}

\usepackage[compact]{titlesec}  
\titlespacing{\section}{0pt}{12pt}{7pt}
\titlespacing{\subsection}{0pt}{9pt}{4pt}

\newcommand{\frb}{FRB~20201124A}

\author[Main et al.]{R.~A.~Main$^1\thanks{Email: \href{mailto:ramain@mpifr-bonn.mpg.de}{ramain@mpifr-bonn.mpg.de}}$,
S.~Bethapudi$^{1}$,
V.~R.~Marthi$^{2}$, 
M.~L.~Bause$^{1}$,
D.~Z.~Li$^{4}$, 
H.-H.~Lin$^{5,6}$, \newauthor 
L.~G.~Spitler$^{1}$,
R.~S.~Wharton$^{3}$, \\
$^{1}$Max-Planck-Institut f{\"u}r Radioastronomie, Auf dem H{\"u}gel 69, D-53121 Bonn, Germany \\
$^{2}$National Centre for Radio Astrophysics, Tata Institute of Fundamental Research, Post Bag 3, Ganeshkhind, Pune - 411 007, India \\
$^{3}$NASA Postdoctoral Program Fellow, Jet Propulsion Laboratory, California Institute of Technology, Pasadena, CA 91109, USA \\
$^{4}$Cahill Center for Astronomy and Astrophysics, MC 249-17 California Institute of Technology, Pasadena CA 91125, USA \\
$^{5}$Institute of Astronomy and Astrophysics, Academia Sinica, Astronomy-Mathematics Building, No. 1, Sec. 4, Roosevelt Road, Taipei 10617, Taiwan, R.O.C \\
$^{6}$Canadian Institute for Theoretical Astrophysics, 60 St. George Street, Toronto, ON M5S 3H8, Canada \\
}
\title[FRB~20201124A Annual Scintillation]{Modelling Annual Scintillation Velocity Variations of FRB~20201124A}

\date{Accepted XXX. Received YYY; in original form ZZZ}

\pubyear{2022}

\begin{document}
\label{firstpage}
\pagerange{\pageref{firstpage}--\pageref{lastpage}}
\maketitle

\begin{abstract}

Compact radio sources exhibit scintillation, an interference pattern arising from propagation through inhomogeneous plasma, where scintillation patterns encode the relative distances and velocities of the source, scattering material, and Earth.
In previous work, we showed that the scintillation velocity of the repeating fast radio burst FRB20201124A could be measured by correlating burst spectra pairs, with low values of the scintillation velocity and scattering timescale suggesting scattering nearby the Earth at $\sim0.4\,$kpc. 
In this work, we have measured the scintillation velocity at 10 epochs spanning a year, observing an annual variation which strongly implies the screen is within the Milky Way.  Modelling the annual variation with a 1D anisotropic or 2D isotropic screen results in a screen distance $d_{l} = 0.40\pm0.04\,$kpc or $d_{l} = 0.46\pm0.06\,$kpc from Earth respectively, possibly associated with material outside of the Local Bubble or the edge of the Orion-Eridanus Superbubble.
Additional measurements particularly at times of low effective velocity will help probe changes in screen properties, and distinguish between screen models.  Where scintillation of an FRB originates in its host galaxy or local environment, these techniques could be used to detect orbital motion, and probe the FRB's local ionised environment.
\end{abstract}


\section{Introduction}

Fast radio bursts (FRBs) are a powerful probe of intervening plasma.  The total electron content along their line of sight is contained in their dispersion measure, while inhomogeneities in electron density result in scattering and scintillation, often concentrated in regions of high electron density.  Scintillation encodes the time delay of scattered paths, and the relative distances and velocities of the emitting source, Earth, and scattering surface \citep{cordes+98}.

Many repeating FRBs undergo periods of extreme activity (e.g. \citealt{gajjar+18, li+21, nimmo+22b, Atel_newFRB}).
In \citet{main+22} (hereafter M22A), we showed that in sufficiently active FRBs (i.e. such that there are many burst pairs separated by $\sim$ minutes), the scintillation timescale $t_s$ can be measured through the pairwise correlation of bursts.  From this, and the easily measured scintillation bandwidth $\nu_{s}$, one can derive a scintillation velocity.
As an extragalactic source, only the Earth's velocity will matter for scintillation within the Milky Way, while the source's velocity will matter if the screen is in the host galaxy.  

In this paper, we measure the scintillation velocity of \frb{} over a full year, and we model the annual variation to determine the properties of the dominant scattering screen. 
We describe our data in Sec. \ref{sec:data} and our scintillation measurements in Sec. \ref{sec:measurements}. In Sec. \ref{sec:annualmodel} we describe the annual scintillation model, and in Sec. \ref{sec:conclusions} we discuss our conclusions and ramifications for future work.

\section{Data}
\label{sec:data}

M22A focused on a single Effelsberg and uGMRT observation.  We continued observing the source with Effelsberg and the uGMRT, with slightly different observing parameters described here.

FRB20201124A has been precisely localised with VLBI to $\alpha = 05^{\rm h}08^{\rm m}03.^{\rm s}5074 \pm 2.7 \textrm{mas}$, $\delta = +26^{\circ}03^{'}38.^{"}5052 \pm 2.6 \textrm{mas}$ \citep{nimmo+22a}. This corresponds to a sightline below the plane towards the galactic anticentre, at $l=178.13^{\circ}, b=-7.92^{\circ}$, while the ecliptic longitude and latitude are $\lambda=79.03^{\circ}, \beta=3.15^{\circ}$, important for the annual scintillation variations modelled in Section \ref{sec:annualmodel}. 

The first Effelsberg observation on MJD 59313 used the PSRIX system \citep{lazarus+16}, recording baseband in a contiguous 1210--1520\,MHz band \citep{hilmarsson+21b}.  One day later, many Effelsberg bursts were detected as part of the EVN campaign to localise the source \citep{nimmo+22a}.  All subsequent Effelsberg observation were made using the Effelsberg Direct Digitization system, recording baseband of the contiguous 1200--1600\,MHz band.  We restrict ourselves to the 200\,MHz band 1270--1470\,MHz throughout, to avoid sensitivity loss in the bandpass edges.

The first uGMRT observation on MJD 59309 used the beam of the incoherent array (IA) due to the uncertain source position, with 2048 channels across 550-750\,MHz, and $655.36\,\mu$s time resolution \citep{marthi+22}.  From these observations we obtained an arcsecond localisation \citep{wharton21b_atel}, so subsequent observations used the phased array (PA) at this position, and our final data product was PA-IA beam, to greatly remove RFI.  Additionally, these observations used 4096 channels, to better resolve the scintillation.  

Bursts were searched for using PRESTO, as described in \citet{hilmarsson+21b, marthi+22}.  Burst detections were made in 3 periods of heightened activity spanning a year, March$-$May 2021 (MJD 59313$-$59359), September 2021 (MJD 59478), and February$-$March 2022 (MJD 59611$-$59655); we summarise our observing campaign, and burst detections in Table \ref{tab:observations}.

\begin{table}
\begin{center}
\caption{Summary of all observations from which we had sufficient bursts to derive $t_s$. Observations marked with $^*$ were included in \citet{main+22}, with slightly different observing parameters as described in Section \ref{sec:data}, and the observation labelled with $\dagger$ comes from the Effelsberg data from the observation used in the EVN localisation \protect{\citep{nimmo+22a}}. The properties of the detected bursts will be analysed in further work.  }
\label{tab:observations}
\begin{tabular}{ cccccc } 
MJD & T$_{\rm obs}$ & N$_{\rm burst}$ & $\nu_{\rm s}$ & $t_{\rm s}$ & $W$ \\
     & (hr) &        & (MHz)  &  (min)    &  (km/s/$\sqrt{\rm kpc}$)  \\
\hline
 \multicolumn{6}{c}{Effelsberg (1270--1470\,MHz)} \\
\hline
59313$*$ & 4 & 20 & $1.24\pm0.08$ & $21.3\pm1.7$ & $24.5\pm2.1$ \\
59314$\dagger$ & 3 & 13 & $1.24\pm0.08$ & $19.0\pm1.2$ & $27.5\pm2.0$ \\
59478 & 4 & 9 & $1.27\pm0.23$ & $78.3\pm9.6$ & $6.7\pm1.0$ \\
59611 & 3 & 53 & $1.29\pm0.15$ & $20.8\pm1.0$ & $25.5\pm1.9$ \\
59612 & 5 & 49 & $1.29\pm0.15$ & $22.0\pm1.2$ & $24.2\pm1.9$ \\
59624 & 4 & 12 & $1.5\pm0.4$ & $27.0\pm4.6$ & $21.2\pm4.5$ \\
59655 & 2 & 7 & $1.6\pm0.4$ & $39.0\pm10.8$ & $15.4\pm5.0$ \\
\hline
 \multicolumn{6}{c}{uGMRT (550--750\,MHz)} \\
\hline
59310$*$ & 3 & 48 & $0.09\pm0.01$ & $8.4\pm2.4$ & $36.2\pm11.4$ \\
59348 & 2 & 13 & $0.07\pm0.01$ & $6.8\pm1.3$ & $39.8\pm8.4$ \\
59359 & 3 & 38 & $0.07\pm0.01$ & $6.2\pm2.2$ & $43.6\pm17.4$ \\
\hline
\end{tabular}
\end{center}
\end{table}

\section{Scintillation Measurements}
\label{sec:measurements}

Extraction of burst spectra and scintillation measurements are almost identical to M22A, we briefly summarise the steps, and note differences here.

At Effelsberg, we channelise the baseband data with 1024 channels, corresponding to $195.3125\,$kHz, while at uGMRT the channel size is $97.65625$\,kHz for the first and $48.828125$\,kHz for subsequent observations.
First, channels corrupted by RFI were manually identified and masked.
For each burst, the `on' region is identified as a contiguous window where the burst is detected with $>5 \sigma$, and an `off' region $100-50\,$ms preceding each burst. The data were divided by the background from the off-region, subtracting 1, and burst spectra were obtained by averaging in time over the on-window. 
The smooth intrinsic spectrum of each burst is estimated as a Gaussian filter of each spectrum, with kernel size of $BW/8$.
The frequency extent of each burst is determined by the region the smoothed spectrum is above $0.5 \sigma_{I}$, the noise estimated in the off-region. Finally, each measured spectrum is divided by the smoothed spectrum, leaving only the variations caused by scintillation.

\begin{figure}
\centering
\includegraphics[width=0.85\columnwidth, clip=true]{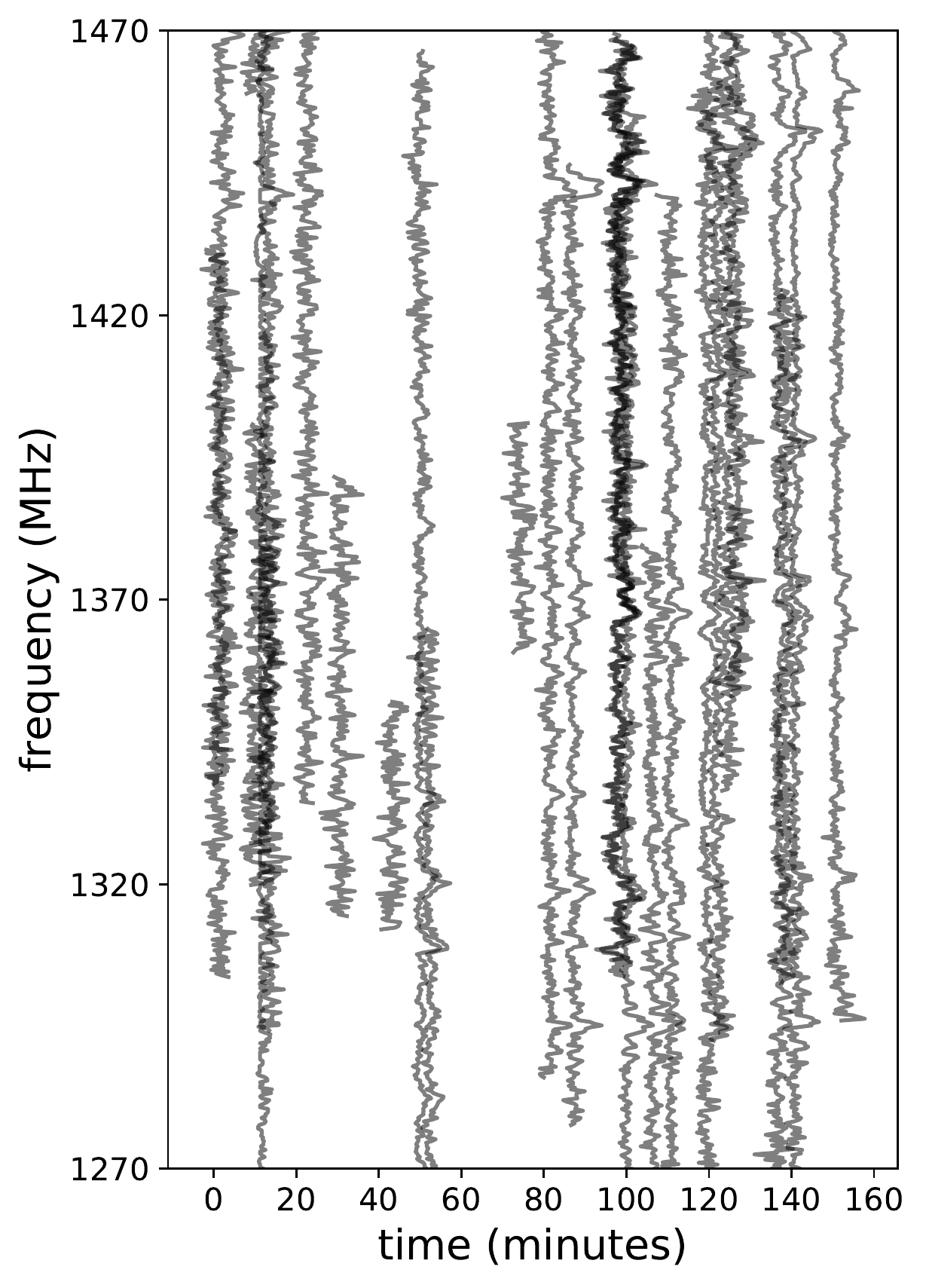}
\vspace{-10pt}
\caption{Dynamic spectrum from MJD 59611, the Effelsberg observation with the largest number of burst detections. The scintillation pattern of nearby bursts are evidently highly correlated. }
\label{fig:dynspec}
\end{figure}

For each observation, we then had $N$ burst scintillation spectra $I(t_i, \nu)$ at discrete times $t_i$ sampling the scintillation pattern unevenly in time (e.g. Fig. \ref{fig:dynspec}).  We correlate each burst pair, giving a measure of $r(\Delta t_{ij}, \Delta \nu)$, with the error on each $r_{ij}$ generated through MCMC method with the off-region noise $\sigma_{I}$ as described in M22A.

A Gaussian is fit to all values of $r(\Delta t_{ij}, \Delta \nu=0)$, and the scintillation timescale $t_s$ is the half width at $1/e$.  We note that contrary to M22A, $t_s$ is fit to the correlation of each burst pair, rather than the ACF binned in time.
To measure the scintillation bandwidth for each observation, we restrict ourselves only to bursts detected over more than half of the band.  To account for frequency evolution, at Effelsberg we split the band into two, and at uGMRT in four, and fit each sub-band separately with a Lorentzian. The scintillation bandwidth $\nu_{s,i}$ in each subband is the half-width at half maximum (HWHM), and the final measurement $\nu_s$ is the weighted sum of $\nu_{s,i} (\nu_{\rm band}/\nu_{\rm ref})^{4}$, and the error is the weighted error on the mean, referenced to 650\,MHz at uGMRT and 1400\,MHz at Effelsberg.  The refractive timescale $t_{r} \approx \frac{2 \nu}{\nu_s} t_{s}$ \citep{rickett90}, which is $\gtrapprox 30\,$days for all of the Effelsberg observations.  As the scintillation bandwidth is not expected to vary significantly below this timescale, we average $\nu_{s}$ for observations on subsequent days (MJDs 59313$-$59314, 59611$-$59612), to reduce noise.

\begin{figure}
\centering
\includegraphics[width=1.0\columnwidth, clip=true]{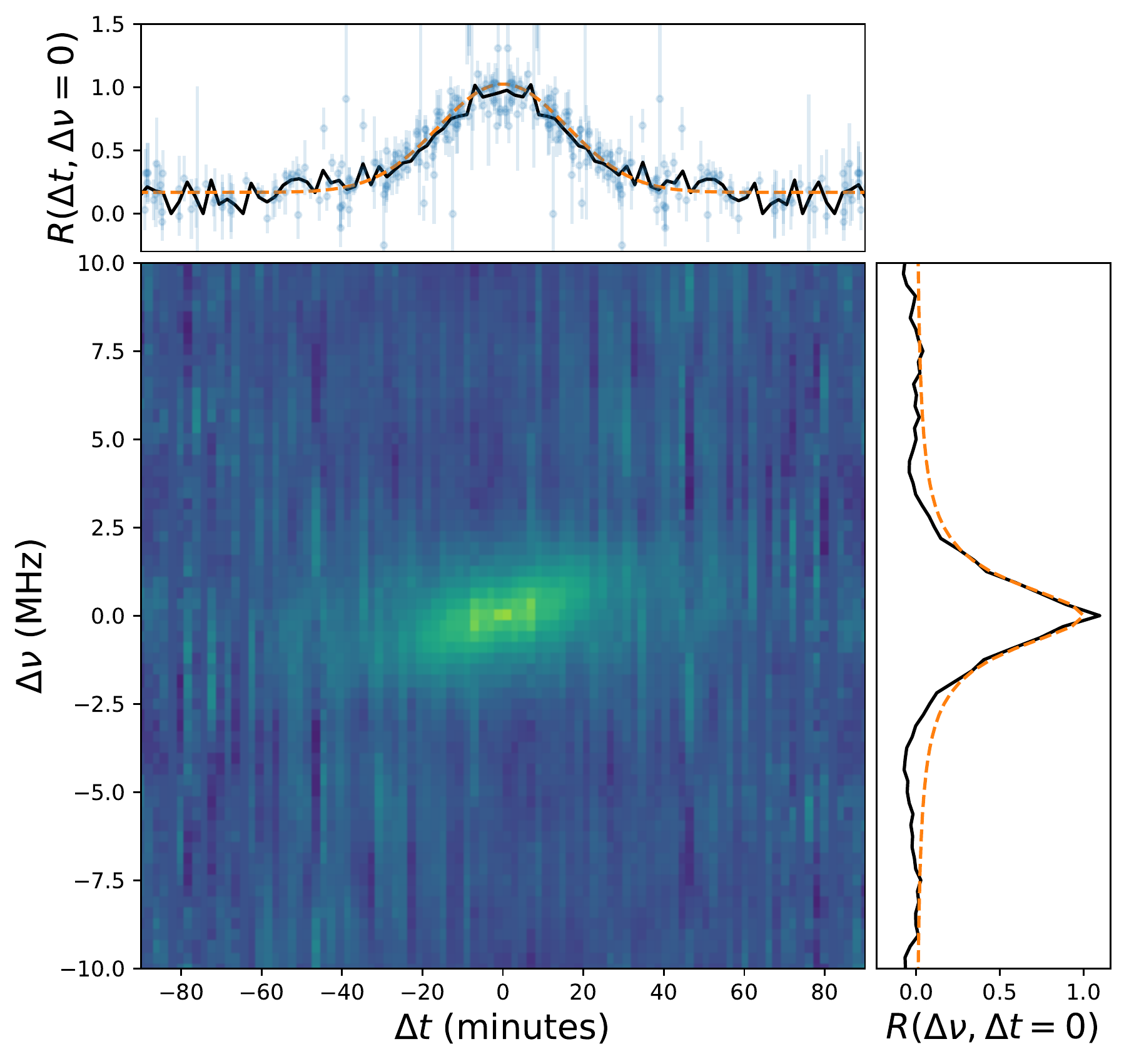}
\hspace{-10pt}
\caption{\textit{Image:} Binned 2D Autocorrelation function of the burst spectra from MJD 59611,  slices through $\Delta \nu=0$ (\textit{top panel}) and $\Delta t=0$ (\textit{right panel}). The orange lines show the model fit for scintillation timescale and bandwidth. We note that the ACF is non-uniformally sampled in time leading to the vertical artefacts; $R(\Delta t, \Delta \nu = 0)$ is fit to the correlation of each burst pair (shown as blue points), rather than the binned ACF.
There is a clear tilt in the ACF suggesting an asymmetric power distribution across the scattering screen (see Sec. \ref{sec:measurements})}
\label{fig:ACF}
\end{figure}

With enough burst pairs, a uniformily sampled 2D autocorrelation function (ACF) $R(\Delta t, \Delta \nu)$ can be constructed through the weighted sum of all burst pairs in bins of $\Delta t$.  We show the ACF from MJD 59611 with 2\,minute bins in Figure \ref{fig:ACF}, along with cuts through $R(\Delta t, \Delta \nu=0)$, $R(\Delta t=0, \Delta \nu)$ and the Gaussian and Lorentzian corresponding the best fit scintillation timescale and bandwidth overplotted.  The ACF cuts are fit well, although there is a clear tilt in the 2D ACF which is unaccounted for.  Such tilts reflect a spatial asymmetry in the scattered power, likely reflecting local phase gradients across the screen \citep{cordes+06, rickett+14}, and suggesting the screen is not isotropic.

As seen in pulsar giant pulses, nano- or micro-shots result in additional intrinsic spectral structure (e.g. \citet{bilous+15}), and would reduce the expected correlation of adjacent pulses to $1/3$ \citep{cordes+04, main+21a}. The spectra of nearby bursts correlate at $\sim 100\,\%$, suggesting there is no intrinsic spectral structure beyond the smooth burst envelope used for normalising the burst spectrum.

\section{Annual Scintillation Model}
\label{sec:annualmodel}

In M22A, we used the expression for scintillation velocity from \citet{cordes+98}, which in the limit of an extragalactic source with $D \gg d_l$ (where $D$ is the source distance, and $d_l$ is the screen distance), the scintillation velocity $V_\mathrm{ISS}$ of a thin screen is related to $\nu_{s}, t_{s}$ as
\begin{equation}
    W \equiv \frac{V_\mathrm{ISS}}{\sqrt{d_l}} \approx 27800 \frac{\sqrt{2 \nu_{s} }}{f t_{s}} ~\textrm{km~s}^{-1}\textrm{kpc}^{-0.5}  \approx \frac{|\textbf{v}_{\earth} - \textbf{v}_{l}|_{||}}{\sqrt{d_l}},
\end{equation}
with reference frequency $f$ in GHz, $\nu_{s}$ in MHz, $t_{s}$ in seconds, and where the scaled effective velocity $W$ was defined for convenience,
separating the unknowns from the measured values.  

\begin{figure}
\centering
\includegraphics[width=1.0\columnwidth, trim=0cm 0cm 0cm 0cm, clip=true]{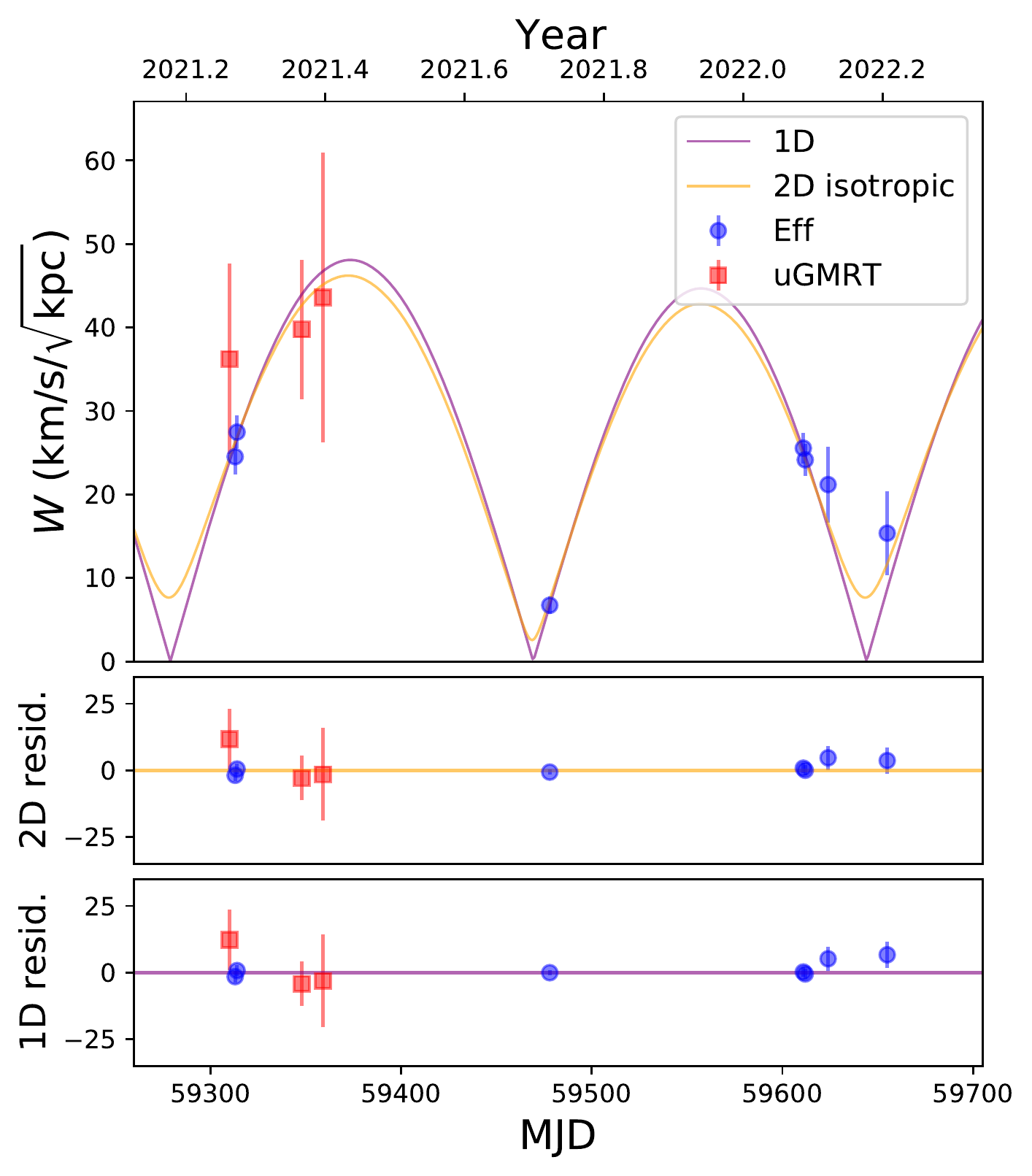}
\vspace{-15pt}
\caption{Measurements of the scintillation velocity, and the best fit 1D and 2D annual models \textit{(top)}, and residuals from the 2D \textit{(top)} and 1D fit \textit{(bottom)}.}
\label{fig:annual}
\end{figure}

\begin{figure}
\centering
\includegraphics[width=0.95\columnwidth, trim=0cm 0cm 0cm 0cm, clip=true]{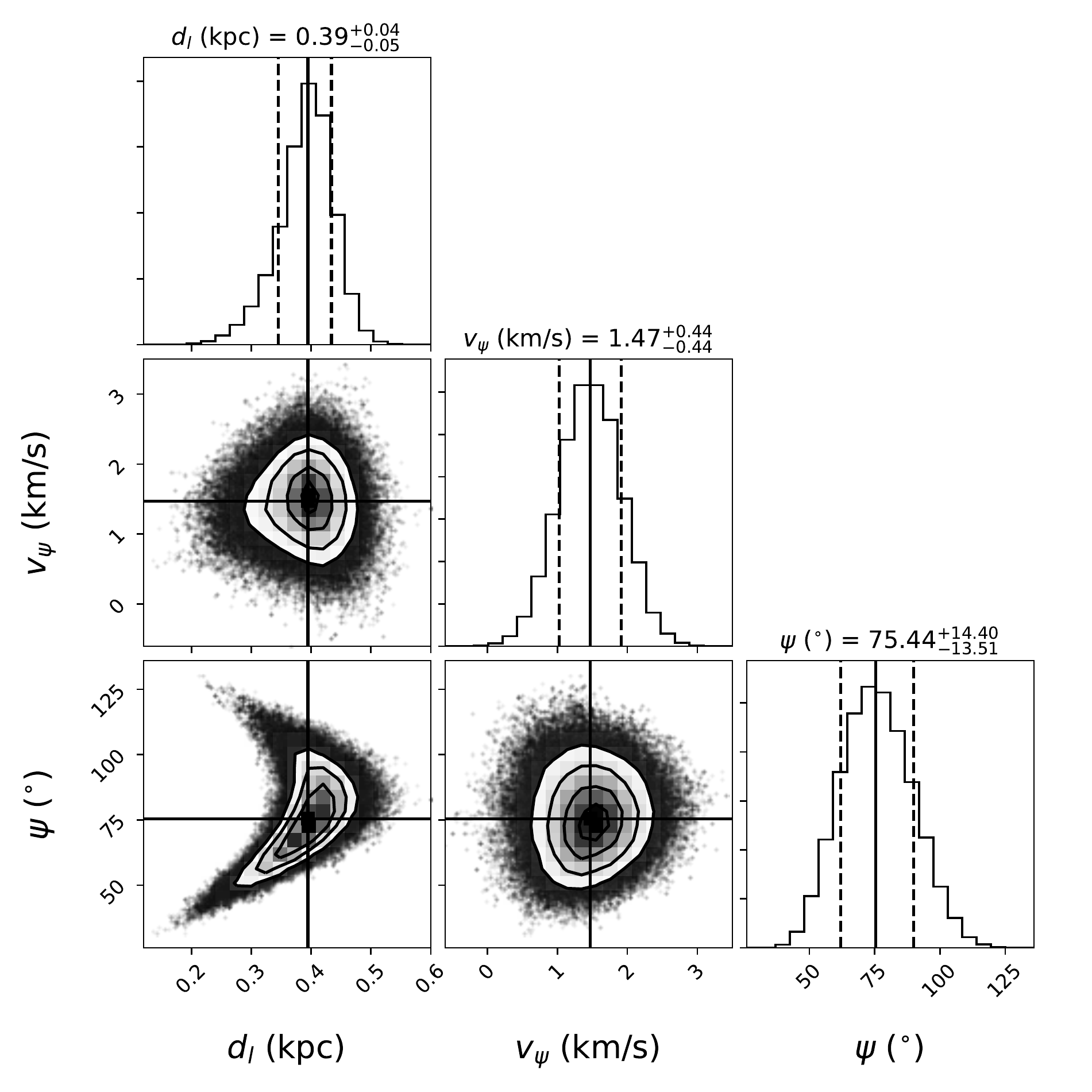} \\
\includegraphics[width=0.95\columnwidth, trim=0cm 0cm 0cm 0cm, clip=true]{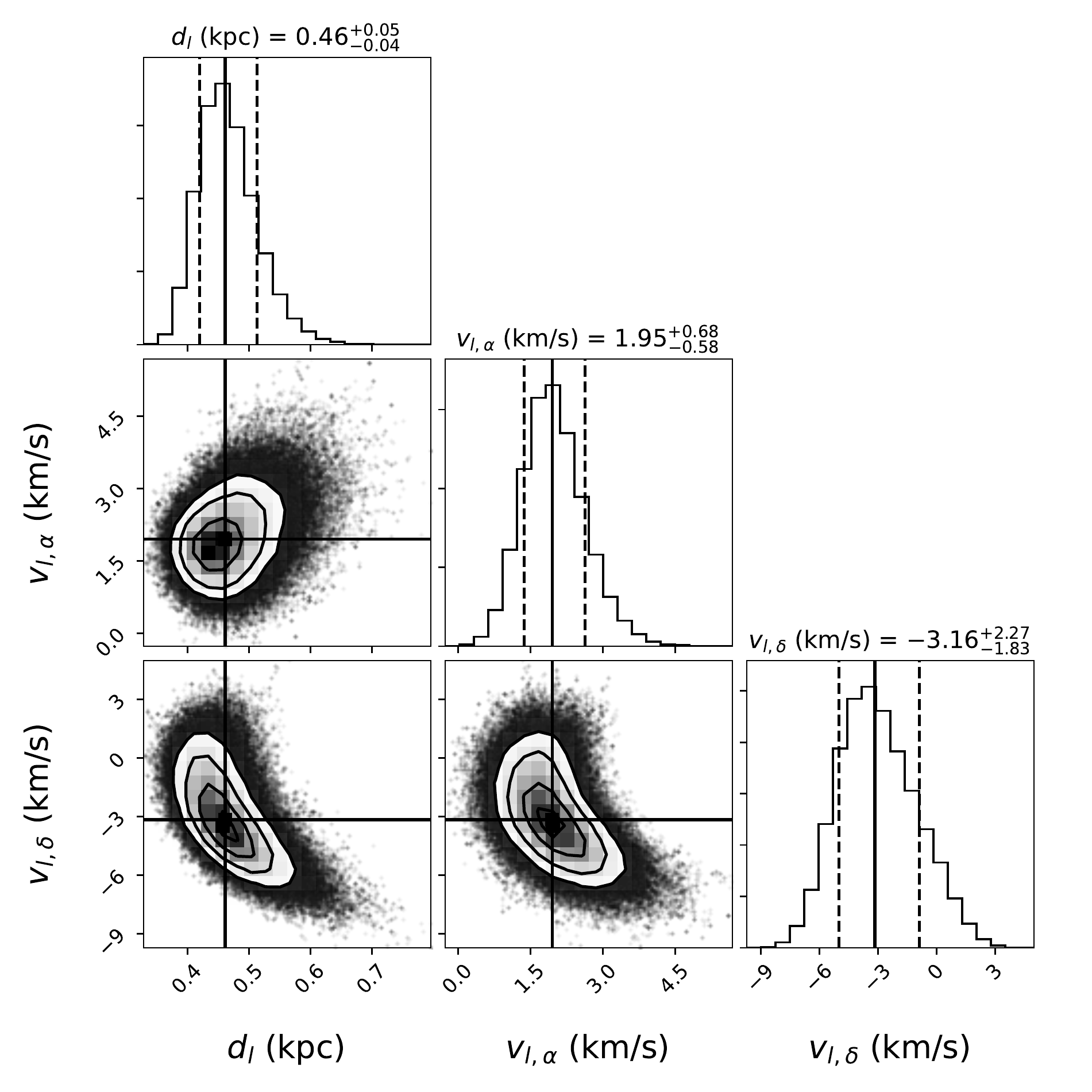}
\vspace{-5pt}
\caption{Posterior plots for the 1D (\textit{top}) and 2D (\textit{bottom}) annual models described in Sec. \ref{sec:annualmodel}, plotted using \textsc{corner.py} \protect{\citep{foremanmackey16}}. The contours show the $68\%$ and $95\%$ confidence intervals. }
\label{fig:posteriors}
\end{figure}

We consider two limiting cases of a fully anisotropic 1D screen, an isotropic 2D screen.
We have full knowledge of $\boldsymbol{\textbf{v}}_{\earth}$, so the only unknowns are the screen distance, and two parameters describing the geometry and velocity of the screen.    
For a 1D screen, we are only sensitive to the velocity parallel to the screen's axis on the sky; the two unknowns are the screen axis $\psi$, and the screen velocity parallel to $\psi$, $v_{l, \psi}$,
\begin{equation}
    W_{\rm 1D} = \frac{1}{\sqrt{d_l}} |v_{\earth, \rm ra} \sin(\psi) + v_{\earth, \rm dec} \cos(\psi) - v_{l, \psi}|
\end{equation}
For a 2D, isotropic screen, the scintillation pattern is sensitive to the total transverse velocity, and the only additional unknown is the 2D velocity of the screen $v_{l,\rm ra}, v_{l,\rm dec}$,
\begin{equation}
    W_{\rm 2D} = \frac{1}{\sqrt{d_l}} \sqrt{ (v_{\earth, \rm ra} - v_{l,\rm ra})^2  + (v_{\earth, \rm dec} - v_{l,\rm dec})^2 }.
\end{equation}
We predict Earth's velocity vector in right ascension and declination towards \frb{} using \textsc{scintools} \citep{scintools}.  We fit our data using a Markov chain Monte Carlo (MCMC), using \textsc{emcee}, with uniform priors on the screen parameters.  The results of our model fitting are shown in Figure \ref{fig:annual}, the posteriors shown in Figure \ref{fig:posteriors}, and best fit parameters are in Table \ref{tab:resultstable}.  

\begin{table}
    \caption{Results of the annual scintillation models, with parameters defined in Sec. \ref{sec:annualmodel}. }
    \label{tab:resultstable}
    \centering
    \begin{tabular}{ccc}
    Parameters & 1D/anisotropic & 2D/isotropic \\
    \hline
$d_l$ (kpc) & $0.40\pm0.04$ & $0.46\pm0.05$ \\
$\psi$ ($^\circ$) & $76\pm13$ & $-$ \\
$v_{\psi}$ (km/s) & $1.4\pm0.4$ & $-$ \\
$v_{l,\rm ra}$ (km/s) & $-$ & $1.9\pm0.6$ \\
$v_{l,\rm dec}$ (km/s) & $-$ & $-3.1\pm2.1$ \\
Reduced $\chi^2$ & 0.75 & 0.6  \\
    \hline
     \end{tabular}
     \newline
\end{table}

\begin{figure}
\centering
\includegraphics[width=1.0\columnwidth, trim=0cm 0cm 0cm 0cm, clip=true]{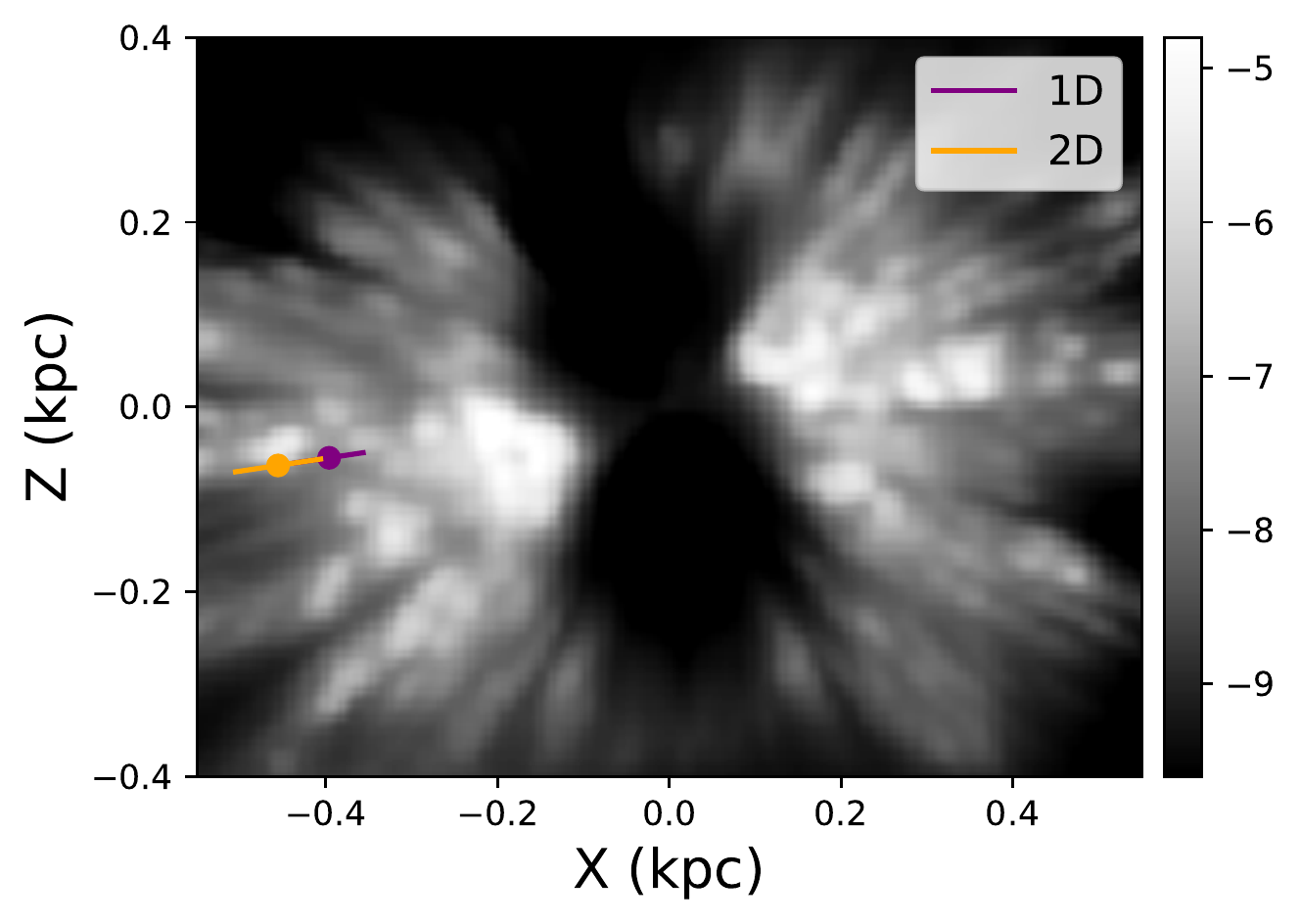}
\vspace{-15pt}
\caption{Screen constraints overlaid over an XZ slice through the local differential extinction map of \protect{\citet{lallement+19}}, where the blue and red points show the screen distance constraints from our annual 1D and 2D models, respectively.  The colorbar show the logarithmic differential extinction in magnitude/pc, and positive values of $X$ point towards the galactic centre, and $Z$ angles above and below the galactic plane. \frb{} is only $1.88^{\circ}$ of longitude away from the Galactic anticentre, motivating the choice of overplotting with the Y=0 slice. }
\label{fig:assoc}
\end{figure}  

\subsection{Screen Association}
\label{sec:screen}

Both the 1D and 2D model fit the data well, with reduced $\chi^2$ of $0.75$ and $0.6$ respectively. 
While the 2D isotropic model is slightly preferred, The ACFs are tilted, suggesting local DM slopes and screen anisotropies which are not accounted for; a general anisotropic screen model with screen axis and axial ratio could also be used, but the reduced $\chi^2$ values less than 1 already suggest that the model is overconstrained.
The annual variation conclusively places the screen in the local region of the Milky Way, with screen distances of $d_{l,1D} = 0.40\pm0.04$\,kpc and $d_{l,2D} = 0.46\pm0.06$\,kpc.   
There is significant correlation between $d_l$ and $\psi$. This degeneracy arises form the low ecliptic latitude of $\beta=3.15^{\circ}$, where the main contribution from changing $\psi$ will be to change the amplitude of $W$ variations, while the phase of an annual scintillation curve will be almost unchanged.

The Sun lies within a $\sim 200\,$pc cavity of hot ($\sim 10^6\,$K) and tenuous ($n_e \approx 0.005\,\textrm{cm}^{-3}$) plasma, known as the Local Bubble (for reviews, see \citealt{cox+reynolds87, frisch+dwarkadas17}).  
It has long been suggested that the Local Bubble boundary is a dominant source of scattering for pulsars \citep{phillips+92, bhat+98}, reinvigorated in recent times as precise screen distances are measured through annual variations of scintillation arcs \citep{mall+22, sprenger+22, mckee+22, stinebring+22}.

To investigate this connection for FRB20201124A, we compare our screen constraints to a map of the Local Bubble in \citet{lallement+19} derived from GAIA and 2MASS dust extinction (Fig. \ref{fig:assoc}). \citet{pelgrims+20} use this same map to model the inner edge of the Local Bubble boundary, and in \citet{zucker+22} it is found that the majority of star forming complexes lie on, or just beyond this boundary, providing a natural location for dense ionised gas.  

Towards \frb{}, the inner edge of the Local Bubble boundary from this model is at a distance of 156\,pc.  
This is inconsistent with our derived screen distances, 
although the Local Bubble is not a sharp edge, as can be seen from the clear overdensity extending from $\sim 150-400\,$pc along this sightline.

This sightline also passes near the edge of the Orion-Eridanus superbubble, a large $\sim 20^\circ \times 45^\circ$ expanding shell at a distance $\sim 400\,$pc, ionised by UV radiation, and driven by supernovae and stellar winds from the Orion OB association \citep{ochsendorf+15, tahani+22}. This would also provide a natural location for plasma overdensities, and the distance roughly matches our screen estimates.

\section{Conclusions and Future Avenues}
\label{sec:conclusions}

We have modelled annual variation of the scintillation velocity in \frb{}, finding the screen originates in the local region of the Milky Way, potentially associated with ionized material beyond the Local Bubble.
With a higher density of scintillation velocity measurements, particularly at the times of lowest effective velocity where the models are most different, the axial ratio and variable screen velocity could be constrained and the screen's distance measured more precisely.  Unmodeled ACF tilts from phase gradients across the screen will introduce systematic errors in the scintillation measurements, especially if they are changing over time.
In observations with even higher number densities of bursts, the phase gradient could be fitted in the 2D ACF as in \citet{rickett+14}, or measurements could be made using scintillation arcs \citep{stinebring+01}, which 
are a more reliable measure of the relative velocities (e.g. \citealt{reardon+20}).

It is often unknown a priori whether the scintillation screen originates in the Milky Way or the host galaxy of the FRB.  A regular annual curve of scintillation velocity breaks this degeneracy.
The scintillation screen can also be unambiguously constrained to the Milky Way if there is a correspondence between scattering time and angular broadening measured through VLBI \citep{ocker+21}.  With new large-scale VLBI projects coming online, the scattering structures along thousands of sightlines could be constrained; combining all such results will lead to greatly improved models and understanding of the Galactic electron content \citep{cordes+02, yao+17}.

Many repeating FRBs have undergone periods of high activity, sufficient to perform similar scintillation analysis.  These include FRB20121102A and FRB20180989B, both sources with periodic active windows \citep{rajwade+20, cruces+21, chime20_periodicity}, and variable magnetised local environments \citep{michilli+18, mckinven+22}.  FRB20190520B shows large magnetic field reversal \citep{anna-thomas+22, dai+22}, and variable scattering on the timescale of minutes \citep{ocker+23}.
Such periodicities, and variable scattering and RM could be explained by an orbit with a windy companion star \citep{lin+21, li+22}, where an orbital dependence of scintillation velocity would be a smoking gun. 

In cases with scintillation and scattering were present on different scales, it has been argued that scintillation would be quenched unless the two scattering screens obey an inequality such that each screen appears pointlike to the other \citep{masui+15, cordes+19}. However, in PSR B1508+55, \citet{sprenger+22} found evidence of intertwined scintillation from 2 screens; the scintillation pattern of the screen nearby the source is imprinted onto the screen nearby the Earth, providing additional constraints on the geometry and extent of the scattering material.  Incorporating these different propagation constraints will help in elucidating FRB host environments, and determining their progenitors.

\section*{ACKNOWLEDGEMENTS}

RAM thanks Vincent Pelgrims for sharing their model of the Local Bubble surface. 
We thank the PRECISE team for sharing the dynamic spectra from the EVN localisation campaign. 
We thank the staff of the GMRT and Effelsberg who have made these observations possible. The Effelsberg 100-m telescope is operated by the Max-Planck-Institut f{\"u}r Radioastronomie. The GMRT is run by the National Centre for Radio Astrophysics of the Tata Institute of Fundamental Research.  VRM acknowledges the support of the Department of Atomic Energy, Government of India, under project no. 12-R\&D-TFR-5.02-0700. 
RSW is supported by an appointment to the NASA Postdoctoral Program at the Jet Propulsion Laboratory, administered by Oak Ridge Associated Universities under contract with NASA.
Part of this research was carried out at the Jet Propulsion Laboratory, California Institute of Technology, under a contract with the National Aeronautics and Space Administration.

\section*{DATA AVAILABILITY}
The data underlying this article will be shared on reasonable request to the corresponding author.

\bibliography{biblio}{}
\bibliographystyle{mnras}

\bsp	
\label{lastpage}
\end{document}